\documentclass{IEEEcsmag}

\usepackage{hyperref}

\usepackage{upmath}
\usepackage[dvipsnames,table,xcdraw]{xcolor}
\usepackage{color}
\usepackage[most]{tcolorbox}

\newtcolorbox{guidelinebox}[2][]{enhanced,
    sharp corners,
    colback=white,
    colbacktitle=white,
    coltitle=black,
    boxrule=1pt,
    left=5mm,
    right=5mm,
    bottom=2mm,
    top=4mm,
    boxed title style={colframe=white},
    attach boxed title to top center={yshift=-3mm},
    center title,
    title=#2,#1
}

\jvol{XX}
\jnum{XX}
\paper{8}
\jmonth{May/June}
\jname{IT Professional}
\pubyear{2019}

\newcommand*{\img}[1]{%
    \raisebox{-.3\baselineskip}{%
        \includegraphics[
        height=\baselineskip,
        width=\baselineskip,
        keepaspectratio,
        ]{#1}%
   }%
}

\begin{document}

\sptitle{Department: Head}
\editor{Editor: Name, xxxx@email}

\title{Guidelines for Developing Bots for GitHub}

\author{Mairieli Wessel}
\affil{Radboud University, The Netherlands}

\author{Andy Zaidman}
\affil{Delft University of Technology, The Netherlands}

\author{Marco A. Gerosa}
\affil{Northern Arizona University, USA}

\author{Igor Steinmacher}
\affil{Northern Arizona University, USA}

\markboth{Department Head}{{Guidelines for Developing Bots for GitHub}}

\begin{abstract}
Projects on GitHub rely on the automation provided by software development bots to uphold quality and alleviate developers' workload. Nevertheless, the presence of bots can be annoying and disruptive to the community. Backed by multiple studies with practitioners, this paper provides guidelines for developing and maintaining software bots. These guidelines aim to support the development of future and current bots and social coding platforms.
\end{abstract}

\maketitle

\chapterinitial{Bridging the gap} between human collaborative software development and automated processes, bots are used to alleviate the software development workload, improve productivity, and enable use cases for which humans are not realistically suitable~\cite{erlenhov2020empirical}. On social coding platforms, such as GitHub, a bot acts autonomously to some extent, has a user account, and plays a role within the development team, executing tasks that complement the developers' work~\cite{Wessel2020}.

Automating simple, time-consuming, or tedious tasks and collecting dispersed information are some ways that bots support software projects. In previous work, we have found that the adoption of bots helps developers merge more pull requests, and reduces the need for communication between developers~\cite{wessel2020effects}. However, while bots are useful for automating a variety of tasks related to software development, prior research has shown that they have the potential side-effect of disrupting developers in their work~\cite{Wessel2021CSCW}.

\begin{figure*}
\centerline{\includegraphics[width=\linewidth]{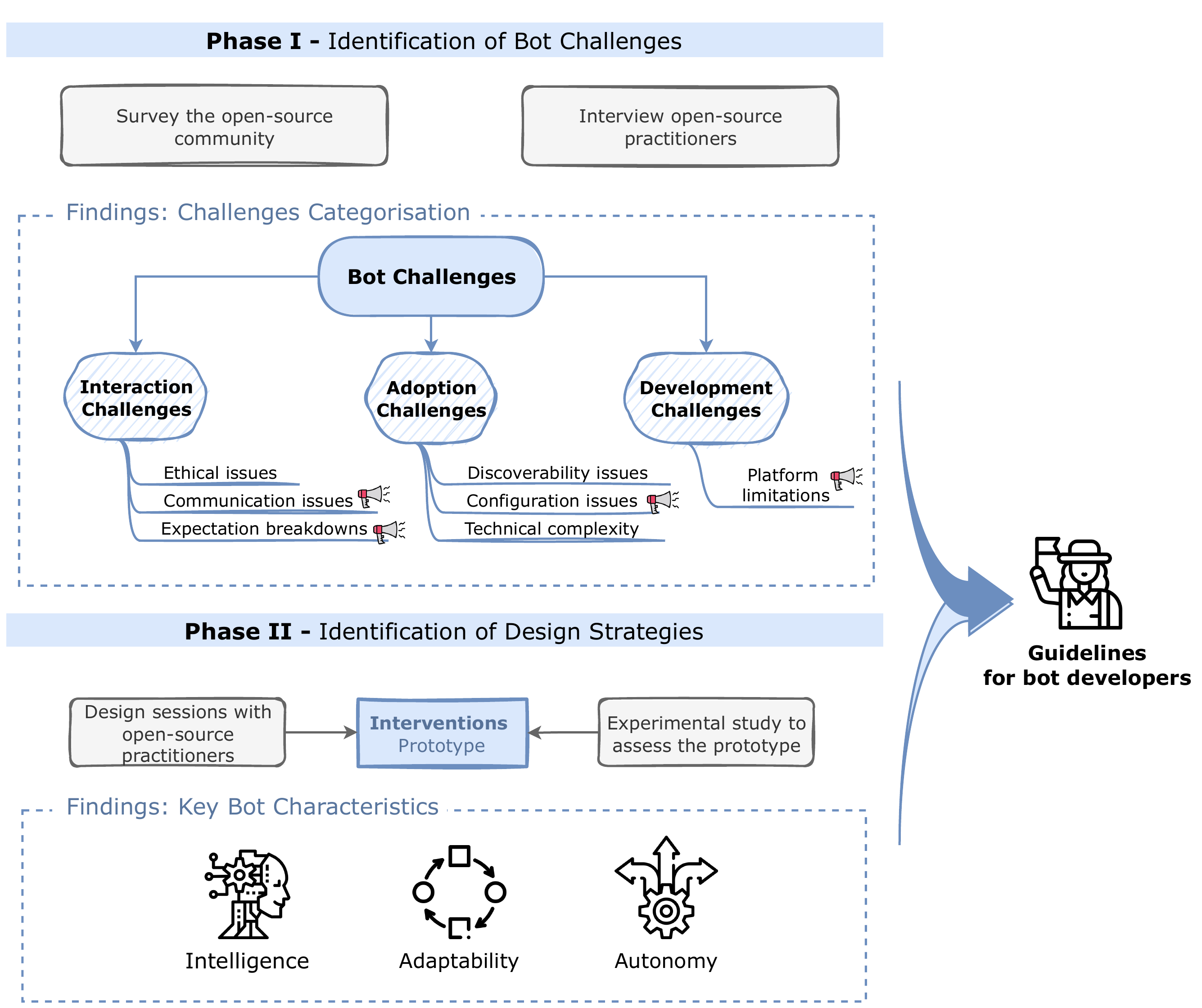}}
\caption{Methodology employed to identify challenges, build a prototype, and create guidelines. The result from Phase I was published at CSCW (\cite{Wessel2018,Wessel2021CSCW}), and Phase II at ICSE 2022~\cite{Wessel2022ICSE}. We added a graphical mark (\img{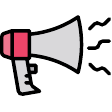}) to identify the challenges related to noisiness, which crosscut the three categories of challenges.}
\label{fig:reserach-design}
\end{figure*}

Surveying and interviewing practitioners, we have found three categories of reported challenges: interaction, adoption, and development challenges~\cite{Wessel2018,Wessel2021CSCW} (see Figure \ref{fig:reserach-design}). Bot noisiness has appeared as a crosscutting concern in all three categories. Noisiness often leads to communication issues and expectation breakdowns. Developers often complain about a bot's verbose messages, timing, and high frequency of actions, which might be caused by platform limitations or bot configuration issues.

Backed by the results of our empirical studies, we have investigated interventions/strategies to mitigate noise and deal with some of the identified challenges~\cite{Wessel2022ICSE}. In line with Erlenhov et al.~\cite{erlenhov2020empirical}'s results, our results indicate that a combination of three different characteristics appears to be relevant for a bot: intelligence, adaptability, and autonomy. Although intelligence and adaptability recurrently appear in the literature as a desired bot characteristic~\cite{erlenhov2020empirical,Lebeuf2019}, they are not yet widely present in bots that work on GitHub~\cite{Wessel2018}.

Backed by the observations gathered from these studies, this paper presents a set of guidelines to help develop software bots for GitHub and (re-)design the human-bot interaction on social coding platforms. We expect that the advances in bot creation frameworks will provide better support for the fulfillment of the guidelines in the future.

\section{Research overview}
\label{sec:challenges}

We have collected evidence of bot noisiness throughout multiple empirical studies as presented in Figure~\ref{fig:reserach-design}. First, we surveyed 205 open source contributors and 23 maintainers~\cite{Wessel2018} and openly asked them about the challenges of using and interacting with bots. To deepen our understanding of these challenges, we interviewed 21 practitioners experienced with bots, including project maintainers, contributors, and bot developers~\cite{Wessel2021CSCW}. The developers' most recurrent complaints are related to annoying bot behaviors. Those behaviors include the case in which bots provide comments with dense information ``\textit{in the middle of the pull request}'', frequently overusing visual elements, and the case in which bots perform repetitive actions, such as creating numerous pull requests and leaving dozens of comments in a row. These behaviors are often perceived as \textit{noise}, which can lead to information and notification overload, which disrupts both human communication and development workflow.

As noise emerged as a central interaction challenge from our empirical analysis, we have further investigated how to overcome it. We created two interventions: (1) a mediator bot that organizes existing bot information in a pull request, and (2) a separate interface for the bot interaction in the pull request~\cite{Wessel2022ICSE}. To design and implement the interventions, we applied Design Fiction~\cite{blythe2014research}, a technique that has been broadly used in the Human-Computer Interaction field to explore and critique future technologies. We presented to 32 open-source maintainers, contributors, bot developers, and bot researchers a fictional story of a mediator bot capable of better supporting developers' interactions on pull requests and operating as a mediator between developers and the existing bots. During synchronous design fiction sessions, participants answered questions to complete the end of the fictional story, discussing the design strategies for the mediator bot and raising concerns about the use of bots. 

Building on the findings of our empirical investigation, we propose a set of guidelines for both bot developers and tool builders. All the guidelines are backed by the evidence previously collected and supported by the literature.

\section{Guidelines for developing bots}
\label{sec:guidelines}

To make bots more effective at accomplishing their tasks, design problems need to be solved to avoid repetitive notifications, provide consistency in the tasks being done, make bots adaptive, and provide clear and contextualized feedback to project contributors and maintainers~\cite{Wessel2021CSCW,Wessel2022ICSE}. To better design the next generation of bots, we provide a set of guidelines along with three main categories: designing bot interaction, facilitating bot adoption, and overcoming platform limitations.

\subsection{Designing bot interaction}
One of the essential aspects of bot interaction is communication. However, the existing bots might fail to provide meaningful information to developers. The most recurrent and central challenge is the introduction of noise into the developers' communication channels. Developers complained about \textit{annoying bot behaviors} such as verbosity, high frequency and timing of actions, and unsolicited actions. Therefore, we present a set of guidelines to support tool builders and bot developers in designing bot interactions.

\begin{guidelinebox}{\textbf{Guideline 1 (G\#1)}}
Provide clear, concise, and well-organized information.
\end{guidelinebox}

\noindent
\textbf{\textit{Interaction challenges:}} We evidenced the need for \textit{background knowledge to interact} with and understand the messages of bots on GitHub. Combined with \textit{lack of context}, it might be extremely difficult for humans to extract meaningful guidance from bots' feedback. In these cases, when a bot message is not clear enough, developers ``\textit{[...] need to go and ask a human for clarity}'', which may generate more work for both contributors and maintainers. 

\noindent\textbf{\textit{What should bot developers do:}}
To reduce the cognitive effort to process bot feedback, it is preferable to prioritize conciseness over completeness. For example, a bot that informs developers whether the changes in a pull request affected the code coverage (i.e., a code coverage bot) should focus on reporting the overall result, and pointing to sources of additional information.

\begin{guidelinebox}{\textbf{Guideline 2 (G\#2)}}
Focus on an appropriate way to show information.
\end{guidelinebox}

\noindent\textbf{\textit{Interaction challenges:}} Another important aspect of bot interaction is the way bots should display information to developers. Developers frequently do not like it when ``\textit{[...] bots put a bunch of information that they try to convey in comments instead of [providing] status hooks or a link somewhere.}'' 

\noindent\textbf{\textit{What should bot developers do:}} Bot developers should identify the best way to convey the information. On GitHub, this can be achieved by exploring possible ways to show information on the platform, which can be either status information or comments. For example, a bot that looks over the code in a pull request and catches quality issues (i.e., a code quality bot) can comment on a pull request to report a list of code formatting issues found. In cases where only an overall status (i.e., passing, falling, blocked) is needed, it is preferable to use status information and avoid overloading pull requests with additional comments.

\begin{guidelinebox}{\textbf{Guideline 3 (G\#3)}}
Provide actionable changes to developers.
\end{guidelinebox}

\noindent\textbf{\textit{Interaction challenges:}} Another recurrent complaint from our survey and interview participants is that bots \textit{do not provide actionable changes} for developers. Some of the messages and outcomes from bots are so strict that they do not guide developers on what they should do next to accomplish their tasks: ``\textit{it is great to see `yes' or `no,' but if it is not actionable, then it is not useful [...]''}.

\noindent\textbf{\textit{What should bot developers do:}} Bot outcomes should be accompanied by actionable and technically sound recommendations by default for the decision-making of developers. For example, a pull request comment from a code coverage bot informing that the coverage decreased is not actionable. However, a comment accompanied by suggested changes is highly actionable because it helps developers to figure out the next steps. 

\begin{guidelinebox}{\textbf{Guideline 4 (G\#4)}}
Avoid overly humanized bot messages.
\end{guidelinebox}

\noindent\textbf{\textit{Interaction challenges:}} Previous studies on human-chatbot interaction have already shown that human users can hold higher expectations with overly humanized bots (e.g., bots that say ``\textit{thank you}'') which can lead to frustration~\cite{gnewuch2017towards}. Our study results underscore that some developers feel uncomfortable interacting with a bot, as mentioned by one participant: `` `\textit{for some people, it is still quite strange, and they are quite surprised by it.}'' Also, receiving ``\textit{thanks}'' from a non-human feels less sincere. 

\noindent\textbf{\textit{What should bot developers do:}} Although developers envision the bot mediator interacting with users through natural language, more direct, and non-humanized bot messages are appreciated. For instance, developers suggested avoiding sentences that do not add to the bot's feedback, such as ``\textit{Hey, I'm here to help you [...]}.''

\begin{guidelinebox}{\textbf{Guideline 5 (G\#5)}}
Make bots' purpose clear.
\end{guidelinebox}

\noindent\textbf{\textit{Interaction challenges:}} By automating and providing feedback on time-consuming tasks (e.g., checking code style or calculating code coverage), bots are intended to reduce the workload of project maintainers and inform project contributors. Nevertheless, maintainers reported that a challenge they see is that ``\textit{contributors don't understand the value of bots for maintainers.}'' We also found that developers with different profiles and backgrounds have different expectations with regard to bot interaction. Bots, for example, \textit{enforce predefined cultural rules} of a community, causing expectation breakdowns for outsiders.

\noindent\textbf{\textit{What should bot developers do:}} It is essential to make the purpose of each bot clear, avoiding expectation breakdowns from both sides. This may be implemented, for example, by including a footnote descriptive sentence or a link to further information about the bot in the bot comment.

\subsection{Facilitating bot adoption}
If maintainers find an appropriate bot, they then have to deal with configuration challenges. Thus, we present advice on how to facilitate the bot adoption process in addition to the guidelines for designing the human-bot interaction. 

\begin{guidelinebox}{\textbf{Guideline 6 (G\#6)}}
Provide options to configure bot notification.
\end{guidelinebox}

\noindent\textbf{\textit{Adoption challenges:}} The study conducted to co-design the mediator bot prototype showed that open-source developers would like to customize aspects of the bot interaction, including notification frequency and timing. Therefore, it is important for bot developers to design a highly customizable bot, providing project maintainers better configuration control over bot actions, rather than just turning off bot comments.

\noindent\textbf{\textit{What should bot developers do:}} In the mediator bot design sessions, developers suggested \textit{scheduling of bot notifications}, so that the bot would avoid notifying developers according to (customizable) timeframes indicating when they do not want interruptions. This may be implemented, for example, using a ``do not disturb'' mode. Another option is \textit{not to notify maintainers until the condition is satisfied}. In this case, the bot would notify the developers only when the predefined conditions are met: ``\textit{I want to be notified about new pull requests after all my tests have passed. And after the bots commented, and if everything is green, then I want to be notified.} The recommendation is that these mechanisms are explicitly announced during bot adoption (e.g., noiseless configuration, preset levels of information).

\begin{guidelinebox}{\textbf{Guideline 7 (G\#7)}}
Include documentation of alternative installation settings to accommodate different types of users.
\end{guidelinebox}

\noindent\textbf{\textit{Adoption challenges:}} It is \textit{difficult to tailor the bot configuration} to fit the needs of a project. Even after maintainers spend the time needed to configure the bot, there is sometimes no way to predict what the bot will do once installed. According to developers' experience, it is ``\textit{easy to install the bot with the basic configuration. However, it is not easy to adjust the configuration to your needs}''.

\noindent\textbf{\textit{What should bot developers do:}} Bot developers should document the bot installation, giving concrete examples of the bot outcomes and possible effects of each configuration choice, and keep it updated. This can also be implemented by creating a FAQ (Frequently Asked Questions) section in a website or repository where the bot code is stored. This is also an opportunity for lowering the entry barrier for new project maintainers, who need to be aware of how each bot works on the project.

\begin{guidelinebox}{\textbf{Guidelines' takeaway}}
Bot developers should envision bots as \emph{socio-technical} rather than \emph{technical} applications, which must be designed to consider human interaction, developers' collaboration, and other ethical concerns.
\end{guidelinebox}

\section{Recommendations to platform builders}
To complement our guidelines, we also explored the platform restrictions since they might limit both the extent of the bots' actions and the way bots communicate. We, therefore, present a set of recommendations for platform builders that would aid bot developers.

\noindent\textbf{Recommendation 1 (R\#1): Enable multiple interaction mechanisms.}

\noindent{\textit{\textit{Platform limitations:}}} Our empirical investigation of bot challenges revealed some limitations imposed by the GitHub platform that restrict the design of bots. As mentioned by one participant: ``\textit{There are still a few things that just cannot be done with the [GitHub] API.}.'' The platform restrictions might limit both the extent of the bots' actions and the way bots communicate.

\noindent{\textit{What should platform builders do:}} It is essential to provide alternative ways for bots to interact on the platform. A developer stated that the platform ideally would provide additional mechanisms since bots interact only through comments. In other environments, such as Slack, developers can interact more flexibly with (chat)bots. On GitHub, this might be achieved by enabling distinct views of the same bot output depending on the developer's role (i.e., maintainer, casual contributor, newcomer) and enabling developers to filter and hide specific bot information.

\noindent\textbf{Recommendation 2 (R\#2): Consider creating a dedicated communication channel for bots.}

\noindent{\textit{\textit{Platform limitations:}}} The interviews we conducted with developers have shown that dealing with bots providing comments with dense information ``\textit{in the middle of the pull request}'' can be ``\textit{[...] a lot more distracting than it is helpful.}'' Bots may overburden developers who already suffer from information overload when communicating online.

\noindent{\textit{\textit{What should platform builders do:}}} To reduce information overload, participants suggested removing bot interactions from the main conversation interface and creating a dedicated place for them. We prototyped this strategy by designing a new tab in the pull request interface; this idea can be leveraged to reshape the interface and better display bot interactions. There is also room for integrating GitHub bots into other developer communication platforms (e.g., Slack, Discord).

\section{The mediator bot}

\begin{figure*}
\centerline{\includegraphics[width=\linewidth]{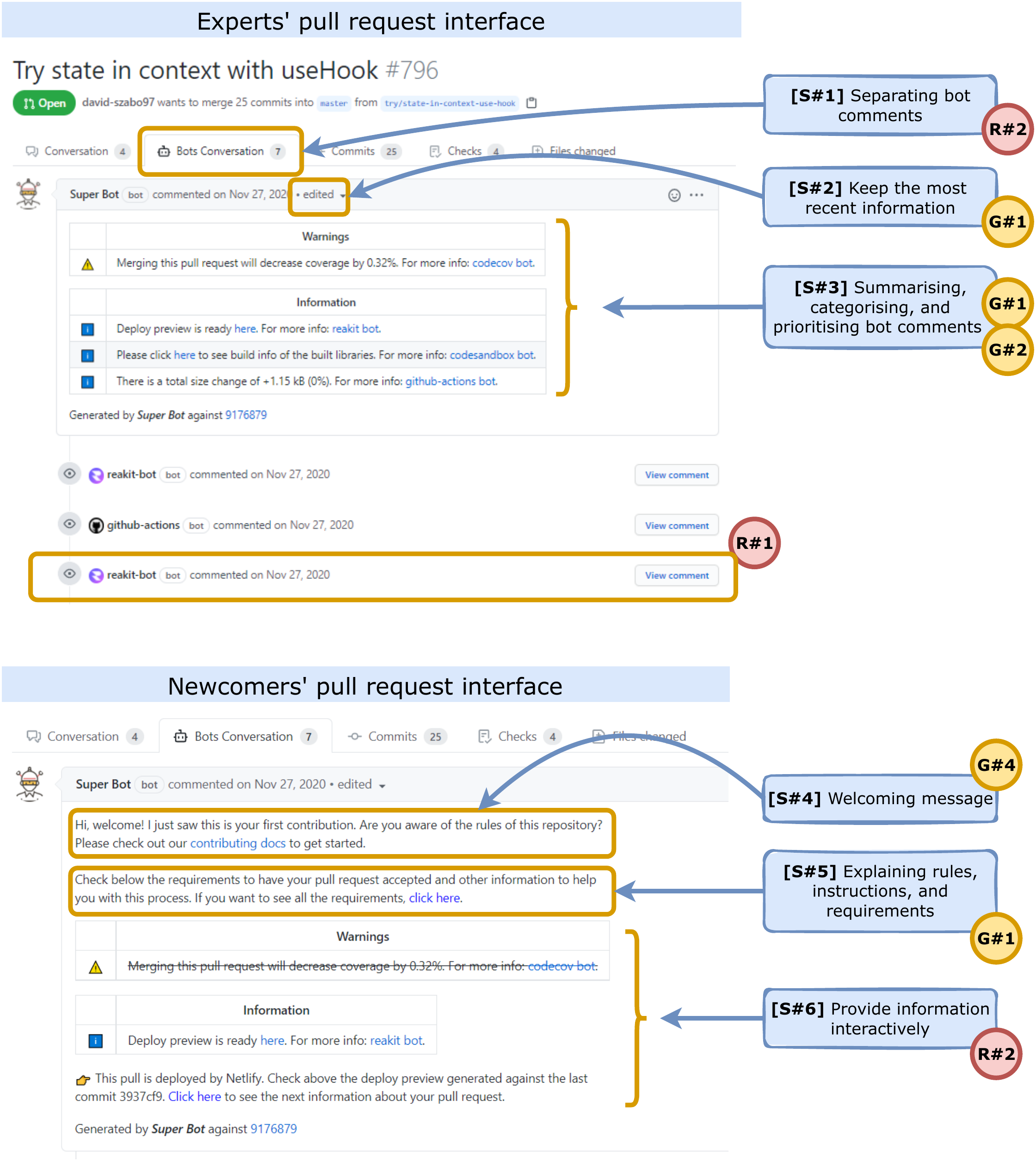}}
\caption{Prototype of the interventions in a real-world scenario on GitHub. It shows the relationship between the design strategies for the mediator bot derived from Phase II (S\#1-6) and our proposed guidelines (G\#1, G\#2, G\#4)/platform recommendations (R\#1, R\#2). The interactive version of the prototype is publicly available on Zenodo~\cite{mairieli_wessel_2021_5675702}.}
\label{fig:prototype}
\end{figure*}
 
To alleviate the concern of bot noisiness in pull requests, we have investigated the concept of a bot that operates as a mediator between developers and the existing bots (i.e., a meta-bot). This section presents our mediator bot prototype and how it connects to the proposed guidelines. Figure~\ref{fig:prototype} provides an overview of the mediator bot design strategies, which we mark throughout the text with (\textit{S\#n}). Firstly, we split our prototype into two different versions: the experts' pull request interface designed to support maintainers and experienced contributors; and the newcomers' pull request interface. We designed a dedicated place for all information and events regarding bots in the pull request (\textit{S\#1}; platform recommendation     \textit{R\#2}). The mediator bot creates a \textit{summary} with the most important information about each bot, then groups them into \textit{categories} (e.g., ``warnings'', ``information'') (\textit{S\#3}; \textit{G\#1-2}).

To avoid inflating the pull requests with several comments from the mediator bot, one suggested strategy is to \textit{keep the most recent information} (\textit{S\#2}; \textit{G\#1}). We include the latest information from each bot in the summary. Reakit bot, for example, posted two comments in the timeline of bot events; however, only one entry is displayed in the summarized table for that bot. In addition, in the timeline of bot events, it is possible to expand all bot comments to see the complete messages (platform recommendation \textit{R\#1}).

In the newcomer's interface, we added a text-based message to fulfill the requirement of welcoming newcomers (\textit{S\#4}; \textit{G\#4}). Beyond presenting a welcoming message, the mediator bot also points the contributor to other sources that can contain information about rules, instructions, and requirements (\textit{S\#5}; \textit{G\#1}) of the project. Thus, we included a link to Reakit’s contributing guidelines.

Another important distinction between the two versions is how the mediator bot displays the information for newcomers versus experts (platform recommendation \textit{R\#1}). We implemented an interactive process of displaying bots' information (\textit{S\#6}). The mediator bot guide newcomers by showing the information from other bots ``\textit{step by step}.'' Study participants deemed this strategy a potential solution to reduce the impact of receiving several different bot notifications simultaneously. As part of this guidance, the mediator bot also refers to contribution guidelines to assist newcomers and present a concise and direct welcoming message.

\section{Conclusion}

Motivated by the growing importance of software bots that act upon the pull-based development model, we have proposed guidelines on how to improve the next generation of bots, considering interaction, adoption, and development challenges identified in prior work. These guidelines can serve bot developers and contributors and maintainers of GitHub projects that use bots in two dimensions: understanding how bots are perceived and how they can be leveraged to support development tasks. Orthogonal to the guidelines, we have also explored the concept of a mediator bot, to alleviate the growing concern of noisiness among bot users. We envision that our guidelines will help developers to produce bots that better automate tasks and further guide developers in collaborative software development.

\section*{Acknowledgments}

This work was partially supported by the Coordenação de Aperfeiçoamento de Pessoal de Nível Superior – Brasil (CAPES) -- Finance Code 001; CNPq grants 141222/2018-2, 314174/2020-6, and 313067/2020-1; the National Science Foundation under Grant numbers 1815503 and 1900903; and the Dutch science foundation NWO through the Vici ``TestShift'' project (No. VI.C.182.032). We also thank the developers who spent their time participating in our research.

\bibliographystyle{IEEEtran}
\bibliography{references}

\begin{IEEEbiography}{Mairieli Wessel}{\,}is an Assistant Professor in the Department of Software Science (SwS) at Radboud University, The Netherlands. She obtained her Ph.D. in Computer Science from the University of São Paulo, Brazil. Her main research interest is in software engineering (SE) and computer-supported cooperative work (CSCW), focused on software bots and open-source development. Her research goal is to design intelligent support for developers by leveraging bots’ capabilities. For more information, visit http://www.mairieli.com
\end{IEEEbiography}

\begin{IEEEbiography}{Andy Zaidman,}{\,}is a Full Professor in software engineering at Delft University of Technology, The Netherlands. He received his M.Sc. and Ph.D. degrees in Computer Science from the University of Antwerp, Belgium, in 2002 and 2006, respectively. His main research interests include software evolution, program comprehension, mining software repositories, software quality, and software testing. He is an active member of the research community and involved in the organization of numerous conferences (WCRE’08, WCRE’09, VISSOFT’14 and MSR’18). In 2013 he was the laureate of a prestigious Vidi midcareer grant, while in 2019 he received the most prestigious Vici career grant from the Dutch science foundation NWO.
\end{IEEEbiography}

\begin{IEEEbiography}{Marco A. Gerosa,}{\,}is a Full Professor at the Northern Arizona University, USA, and a Ph.D. advisor at the University of São Paulo, Brazil. He received his Ph.D. from the Catholic University of Rio de Janeiro in 2006. He researches Software Engineering and CSCW, and recent projects include the development of tools and strategies to support developers' onboarding to open-source software communities and the design of bots and chatbots. He published more than 200 papers and serves on the program committee (PC) of top-tier conferences, such as ICSE, FSE, MSR, ICSME, and CSCW. For more information, visit http://www.marcoagerosa.com
\end{IEEEbiography}

\begin{IEEEbiography}{Igor Steinmacher,}{\,}is an Assistant Professor in the School of Informatics, Computing, and Cyber Systems at the Northern Arizona University (NAU), and was previously at the Federal University of Technology Paraná (UTFPR), Brazil. He received a Ph.D. in Computer Science from the University of São Paulo (USP — Brazil). He researches the intersections of Software Engineering (SE) and Computer Supported Cooperative Work (CSCW). Currently, his research focuses on the behavior of developers in Open Source Communities, including the support of newcomers, the impact of Bots in the community, and gender bias in Open Source Software. His interests include Open Source Software, Human Aspects of Software Engineering, Empirical Software Engineering, and Mining Software Repositories techniques. 
\end{IEEEbiography}
\end{document}